\documentclass[showpacs,showkeys,preprintnumbers, amsmath,amssymb,preprint]{revtex4-1}
\usepackage{amssymb}
\usepackage{amsfonts}
\usepackage{amsmath}
\usepackage{graphicx}
\usepackage{dcolumn}
\usepackage{bm}
\usepackage{makeidx}
\usepackage[latin1]{inputenc}
\usepackage[dvips]{hyperref}

\usepackage{graphicx}
\def\be{\begin{equation}}
 \def\ee{\end{equation}}
 \def\bea{\begin{eqnarray}}
 \def\eea{\end{eqnarray}}

\def\PR#1{{Phys.\ Rev.\ D \bf #1}}
\def\PRL#1{{Phys.\ Rev.\ Lett.\ \bf #1}}

\makeindex

\begin{document}

\title{Motion of particles on a Four-Dimensional Asymptotically AdS Black Hole with Scalar Hair}
\author{P. A. Gonz\'{a}lez}
\email{\,\,\,\,pablo.gonzalez@udp.cl} \affiliation{Facultad de
Ingenier\'{i}a, Universidad Diego Portales, Avenida Ej\'{e}rcito
Libertador 441, Casilla 298-V, Santiago, Chile.}
\author{ Marco Olivares }
\email{marco.olivaresrubilar@gmail.com}
\affiliation{Instituto de F\'{\i}sica, Pontificia Universidad 
Cat\'{o}lica de Valpara\'{\i}so, Avenida Universidad 330, Curauma,\\
Valpara\'{\i}so, Chile.}
\author{Yerko V\'{a}squez}
\email{yvasquez@userena.cl}
\affiliation{Departamento de F\'{\i}sica, Facultad de Ciencias, Universidad de La Serena,\\ 
Avenida Cisternas 1200, La Serena, Chile.}
\date{\today}

\begin{abstract}
Motivated by black hole solutions with matter fields outside their horizon, we study the effect of these matter fields in the motion 
of massless and massive particles. We consider as background a four-dimensional asymptotically 
AdS black hole with scalar hair. The geodesics are studied numerically and we discuss 
about the differences in the motion of particles between the four-dimensional 
asymptotically AdS black holes with scalar hair and their no-hair limit, that is, 
Schwarzschild AdS black holes. Mainly, we found that there are bounded 
orbits like planetary orbits in this background. However, the periods associated to circular orbits 
are modified by the presence of the scalar hair. Besides, we found that some classical 
tests such as perihelion precession, deflection of light and gravitational time 
delay have the standard value of general relativity plus a correction term coming 
from the cosmological constant and the scalar hair. Finally, we found a specific value of the parameter associated to the scalar hair, in order to explain the discrepancy between the theory and the observations, for the perihelion precession of Mercury and light deflection.
\end{abstract}

\maketitle


\tableofcontents


\section{Introduction}

Hairy black holes are interesting solutions of Einstein's Theory
of Gravity and also of certain types of Modified Gravity Theories. The first attempts to couple a scalar field to gravity was done in
an asymptotically flat spacetime finding hairy black hole solutions \cite{BBMB} but these
solutions 
violated the no-hair theorems because they were not physically
acceptable as the scalar field was divergent on the horizon and
stability analysis showed that they were unstable
\cite{bronnikov}.
Then, by introducing a cosmological constant  hairy black hole solutions with a minimally
coupled scalar field and a self-interaction potential in
asymptotically dS space were found, but it
was unstable \cite{Zloshchastiev:2004ny, Torii:1998ir}. Also, a hairy
black hole configuration was reported  for scalar field non-minimally coupled to gravity \cite{Martinez:2002ru}, but
perturbation analysis showed the instability of the solution
\cite{Harper:2003wt,Dotti:2007cp}.
 In the case of a
negative cosmological constant, stable solutions were found
numerically for spherical geometries \cite{Torii:2001pg,
Winstanley:2002jt} and an exact solution in asymptotically AdS
space with hyperbolic geometry was presented in
\cite{Martinez:2004nb} and generalized later to include electric charge
\cite{Martinez:2005di}. Then, 
a generalization to
non-conformal solutions was discussed in \cite{Kolyvaris:2009pc}.
Further hairy solutions in the presence of a cosmological constant
were reported in
\cite{Anabalon:2012ta,Anabalon:2012ih,Bardoux:2012tr, Gonzalez:2013aca, Gonzalez:2014tga}
with various properties. On the other hand, by introducing
a coupling of a scalar field to
Einstein tensor
that acts as an effective
cosmological constant \cite{Amendola:1993uh,Sushkov:2009hk} a hairy black hole solution was presented \cite{Kolyvaris:2011fk}   
Additionally, spherically symmetric hairy black hole
solutions with scalar hair were found 
\cite{Kolyvaris:2013zfa}. 
Additionally, there are also very
interesting recent developments in Observational Astronomy. High
precision astronomical observations of the supermassive black
holes may pave the way to  experimentally test the no-hair
conjecture
 \cite{Sadeghian:2011ub}. Also, there are numerical investigations
 of single and binary black holes in the presence of scalar fields
\cite{Berti:2013gfa}. 

On the other hand, the recent developments in string theory and
specially the application of the AdS/CFT principle to condense
matter phenomena like superconductivity (for a review see
\cite{Hartnoll:2009sz}), triggered the interest of further study
of the behavior of matter fields outside the black hole horizon
\cite{Gubser:2005ih,Gubser:2008px}. In this context, the gauge/gravity duality is a principle which relates strongly
coupled systems to their weak coupled gravity duals. One of
the most well studied system 
is the holographic superconductor. In its simplest form,
the gravity sector is a gravitating system with a cosmological
constant, a gauge field and a charged scalar field with a
potential (for a review see \cite{Horowitz:2010gk}). The dynamics
of the system defines a critical temperature above which the
system finds itself in its normal phase and the scalar field does
not have any dynamics. Below the critical temperature the system
undergoes a phase transition to a new configuration. From the
gravity side this is interpretated as the black hole to acquire
hair while from boundary conformal field theory site this is
interpretated as a condensation of the scalar field and the
system enters a superconducting phase. 

It is known that all solar system observations, such as light deflection, the perihelion shift of planets, the gravitational time-delay among other are described within Einstein's General Relativity. The study of geodesics has been performed under several black hole geometries. For instance, see \cite{Cruz:2004ts, Vasudevan:2005js, Hackmann:2008zz, Hackmann:2008zza, Olivares:2011xb, Cruz:2011yr,  Larranaga:2011fp, Villanueva:2013zta} for the motion of particles on AdS space-time. In this work, motivated by black hole solutions with matter fields outside their horizon, we study their effect in the motion of massless and massive particles in the background of a four-dimensional asymptotically AdS black hole with scalar hair \cite{Gonzalez:2013aca}.   These hairy black holes solutions are characterized by a self-interacting potential that asymptotically tends to the cosmological constant, and the scalar
field is regular everywhere outside the event horizon and null at spatial infinity. The geodesics are studied numerically and we discuss about the differences in the motion of particles between the four-dimensional asymptotically AdS black holes with scalar hair and their no-hair limit, that is, Schwarzschild AdS black holes. Also, we study classical tests such as perihelion precession, deflection of light and gravitational time delay in order to determine the contribution that arises from the scalar hair.

The paper is organized as follows. In Section II we give a brief review of the four-dimensional asymptotically AdS black holes with scalar hair that we will consider as background. In Section III we study the motion of massless and massive particles, and we perform some classical tests such as perihelion precession, deflection of light and gravitational time delay. Finally in Section IV
we conclude.

\section{Four-Dimensional Asymptotically AdS Black Holes with Scalar Hair}
The hairy black hole that we consider is solution of the Einstein-Hilbert action with  a negative cosmological constant and a neutral
scalar field minimally coupled to the curvature having a
self-interacting potential $ V(\phi)$ \cite{Gonzalez:2013aca}. The action is given by
\begin{eqnarray} \label{action}
 S=\int d^{4}x\sqrt{-g}\left(\frac{1}{2 \kappa }R-\frac{1}{2}g^{\mu\nu}\nabla_{\mu}\phi\nabla_{\nu}\phi-V(\phi)\right)~,
 \end{eqnarray}
being the self-interacting potential
\begin{equation}
V\left( \phi \right) =-F\left( 2+\cosh \left( \sqrt{2}\phi \right) \right) +%
\frac{G}{\nu ^{3}}\left( 6\sinh \left( \sqrt{2}\phi \right) -2\sqrt{2}\phi
\left( 2+\cosh \left( \sqrt{2}\phi \right) \right) \right)~.
\end{equation}
Here, the cosmological constant is incorporated in the potential, that is, $
\Lambda=V(0)$ ($V(0)<0$). Where,
$\Lambda=-6l^{-2}/\kappa$, being $l$ the length of the AdS
space and $\kappa=8 \pi G_N$, with $G_N$ the Newton constant. This potential has a global maximum at $\phi =0$.  The equations of motion are
 \begin{eqnarray}
 R_{\mu\nu}-\frac{1}{2}g_{\mu\nu}R=\kappa T^{(\phi)}_{\mu\nu}\label{field1}~,
 \end{eqnarray}
where the energy momentum tensor $T^{(\phi)}_{\mu\nu}$ for the
scalar field is
 \begin{eqnarray}
 T^{(\phi)}_{\mu\nu}=\nabla_{\mu}\phi\nabla_{\nu}\phi-
 g_{\mu\nu}[\frac{1}{2}g^{\rho\sigma}\nabla_{\rho}\phi\nabla_{\sigma}\phi+V(\phi)]\label{energymomentum}~,
 \end{eqnarray}
and the Klein-Gordon equation of the scalar field 
reads \be \Box \phi =\frac{d V}{d \phi}~. \label{klg} \ee
The following metric is solution of the theory defined by (\ref{action})
 \begin{eqnarray}
 ds^{2}=-f(r)dt^{2}+f^{-1}(r)dr^{2}+a^{2}(r)d \sigma_{k}^2~, \label{metricBH}
 \end{eqnarray}
where
 \begin{equation}
f\left( r\right) =k+Fr\left( r+\nu \right) +\frac{G}{\nu
^{3}}\left( -\nu
\left( \nu +2r\right) +2r\left( r+\nu \right) \ln \left( \frac{r+\nu }{r}%
\right) \right) ~,
\end{equation}
\begin{eqnarray}
 a^{2}(r)=r(r+\nu)~,
 \label{ar}
 \end{eqnarray}
and the scalar field is
\begin{equation}
\phi \left( r\right) =\frac{1}{\sqrt{2}}\ln \left( 1+\frac{\nu }{r}\right) ~.
\label{field}
\end{equation}
In expression (\ref{metricBH}), $d \sigma_{k} ^2$ is the metric of the spatial 2-section, which
can have positive, negative or zero  curvature, and the coordinates are defined in the ranges
$0<r<\infty$, $-\infty<t<\infty$, $0\leq \theta <\pi$, $0\leq \phi <2\pi$. For the lapsus function 
 $k=1,0,-1$ parametrizes the curvature of the spatial 2-sections and $F$, $G$ are  constants being proportional to
the cosmological constant and to the mass respectively. It was shown, that for spherical horizons $k=1$ there is no phase transition of the hairy
asymptotically AdS  black holes to Schwarzschild AdS
black hole. However, for hyperbolic horizons $k=-1$ there exists a phase transition only for
negative masses, and the hairy black hole dominates for small
temperatures, while for large temperatures the topological black
hole would be preferred, for more details see \cite{Gonzalez:2013aca}.

\begin{figure}[h]
\begin{center}
\includegraphics[width=0.5\textwidth]{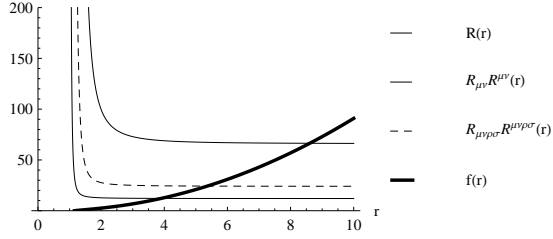}
\end{center}
\caption{The behavior of $R(r)$, $R^{\mu\nu}R_{\mu\nu}(r)$, $R^{\mu\nu\lambda\tau}R_{\mu\nu\lambda\tau}(r)$ and $f(r)$ with $k=1$, $\protect%
\nu =-1$, $F=1$ and $G=2$.} \label{invariantes}
\end{figure}

In the next section we perform a numerical analysis of the geodesics by considering the hairy black hole solution. So, without loss of generality, we consider the following values for the parameters: $k=1$, $\nu =-1$, $F=1$ and $G=2$. Thus, in order to show that these parameters yield a hairy black hole solution we plot in 
Fig. \ref{invariantes} the  behavior of the metric function
$f\left( r\right)$ that changes sign for $r=1.15$ signaling the
presence of an horizon. Also we plot the
behavior of the Ricci scalar $R(r)$, the principal quadratic invariant of the Ricci tensor $R^{\mu\nu}R_{\mu\nu}(r)$, and the Kretschmann scalar
$R^{\mu\nu\lambda\tau}R_{\mu\nu\lambda\tau}(r)$ and we observe that there is not
Riemann curvature singularity outside the horizon. Also, we observe that the Riemann curvature singularities are covered by the horizon. Therefore, the choice of parameters mentioned above gives a hairy black hole solution which is asymptotically AdS. 

\section{Geodesics}
\label{STL}
In order to find the geodesics of the space-time described by (\ref{metricBH}), we will solve the Euler-Lagrange equations for the variational problem associated with this metric. 
The Lagrangian associated to the metric
(\ref{metricBH}) is given by
\begin{equation}\label{tl4}
  2\mathcal{L}=- f(r)\dot{t}^2+
  \frac{\dot{r}^2}{f(r)}+a^2(r)(\dot{\theta}^2+\sin^2\theta\,\dot{\phi}^2)=-m~,
\end{equation}
where $\dot{q}=dq/d\tau$, and $\tau$ is an affine parameter along the geodesic that
we choose as the proper time. Since the Lagrangian (\ref{tl4}) is
independent of the cyclic coordinates ($t, \phi$), then their
conjugate momenta ($\Pi_t, \Pi_{\phi}$) are conserved and
the equations of motion reads
\be \dot{\Pi}_{q} - \frac{\partial \mathcal{L}}{\partial q} = 0~,
\label{w.10} \ee
where $\Pi_{q} = \partial \mathcal{L}/\partial \dot{q}$
is the conjugate momenta to the coordinate $q$.  The above equation can be written as
\begin{equation}
\dot{\Pi}_{t} =0~, \quad \dot{\Pi}_{r} =-{\dot{t}^{2}\over 2}{df(r)\over dr}+ 
{\dot{r}^{2}\over 2}{df(r)^{-1}\over dr}+a(r){da(r)\over dr}(\dot{\theta}%
^{2}+\sin^{2}\theta \dot\phi^2)~,
\label{w.11a}
\end{equation}
\begin{equation}
\dot{\Pi}_{\theta} = a^2(r)\sin\theta \cos\theta \,\dot\phi^2, \quad
\textrm{and}\quad \dot{\Pi}_{\phi}=0~,
\label{w.11b}
\end{equation}
which yields
\begin{equation}
\Pi_{t} = -f(r) \dot{t} , \quad \Pi_{r}= {\dot{r}\over f(r)}~,
\label{w.11c}
\end{equation}
\begin{equation}
\Pi_{\theta} = a^{2}\dot{\theta}~ , \quad
\textrm{and}\quad \Pi_{\phi}
= a^{2}\sin^{2}\theta \dot{\phi}~.
\label{w.11d}
\end{equation}
Now, without loss of generality, we consider that the motion is developed in the invariant plane
 $\theta  = \pi/2$ and $\dot\theta =0$, which is characteristic of the central fields. With this choice, Eqs. (\ref{w.11c}) and (\ref{w.11d})  become
 \begin{equation}
\Pi_{t} = -f(r) \dot{t}\equiv -\sqrt E~, \quad \Pi_{\phi}= a^{2}\dot{\phi}\equiv L~,
\label{w.11c}
\end{equation}
where $E$ and $L$ are dimensionless integration constants associated to each of them.
So, inserting
equations (\ref{w.11c}) into equation (\ref{tl4}) we obtain
\begin{equation}\label{tl7}
  \left(\frac{dr}{d\tau}\right)^2=E-V(r)~,
\end{equation}
where $V(r)$ is the effective potential
given by
\begin{equation}\label{tl8}
  V(r)=f(r)\left[m+\frac{L^2}{r(r+\nu)}\right]~,
\end{equation}
where $m$ is the test mass. Finally, by normalization, we shall consider that $m=1$ for massive particles and $m=0$ for photons.  

\subsection{Time like geodesic}
\begin{figure}[h]
\begin{center}
\includegraphics[width=0.5\textwidth]{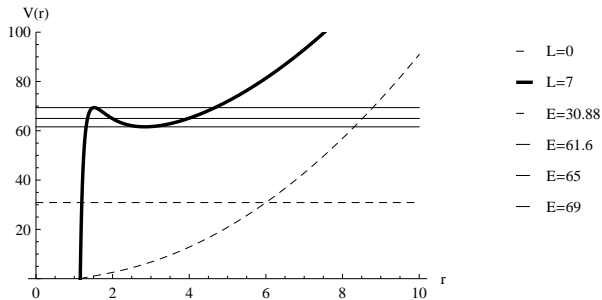}
\end{center}
\caption{The behavior of $V(r)$ for radial ($L=7$) and non-radial ($L=0$) particles, with $k=1$, $\nu =-1$, $F=1$ and $G=2$.} \label{plots1}
\end{figure}

In order to observe the possible orbits, we plot the effective potential for massive particles (\ref{tl8}) which is shown in Fig. (\ref{plots1}). In the following, we describe the radial motion and the angular motion.

\subsubsection{Radial motion}
\begin{figure}[h]
\begin{center}
\includegraphics[width=0.5\textwidth]{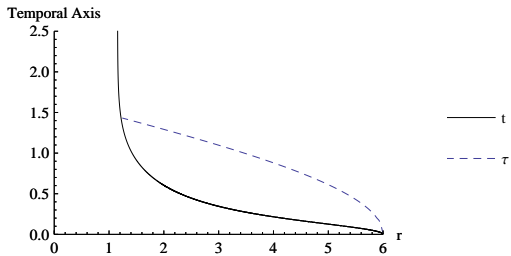}
\end{center}
\caption{The behavior of the proper ($\tau$) and coordinate ($t$) time as function of $r$, with $k=1$, $\nu =-1$, $F=1$ and $G=2$.} \label{plots2}
\end{figure}

In this case $L=0$. The particles always fall into the horizon from an upper distance determined by the constant of motion $E=30.88$. This fact is due to the attractive force generated by the proportional term to the cosmological constant, see Fig. (\ref{plots1}). In Fig. \ref{plots2} we plot the proper  ($\tau$) and coordinate ($t$) time as function of $r$ for a particle falling from a finite distance with zero initial velocity, we can see that the particle falls towards the horizon in a finite proper time. The situation is very different if we consider the trajectory in the coordinate time, where $t$ goes to infinity.     
This physical result is consistent with the Schwarzschild AdS black hole. 

\subsubsection{Angular motion}
For the angular motion we consider $L>0$. The allowed orbits depend on the value of the constant $E$. 
\begin{itemize}
\item
If $E=61.6$ the particle can orbit in a stable
circular orbit at $r_s=2.84$, see Fig. \ref{plots1}.

\item
If $E=69.4$ the particle can orbit in an unstable
circular orbit at $r_u=1.518$. Also, there are two critic orbits that approximates asymptotically to the unstable circular orbit. First kind, the particle starts from the rest and a finite distance greater than the unstable radio, see Fig. \ref{plot3}. The second kind, the particle starts from  a finite distance greater than the horizon, but smaller than the unstable radio, see Fig. \ref{plot3}.
\begin{figure}[h]
\begin{center}
\includegraphics[width=0.35\textwidth]{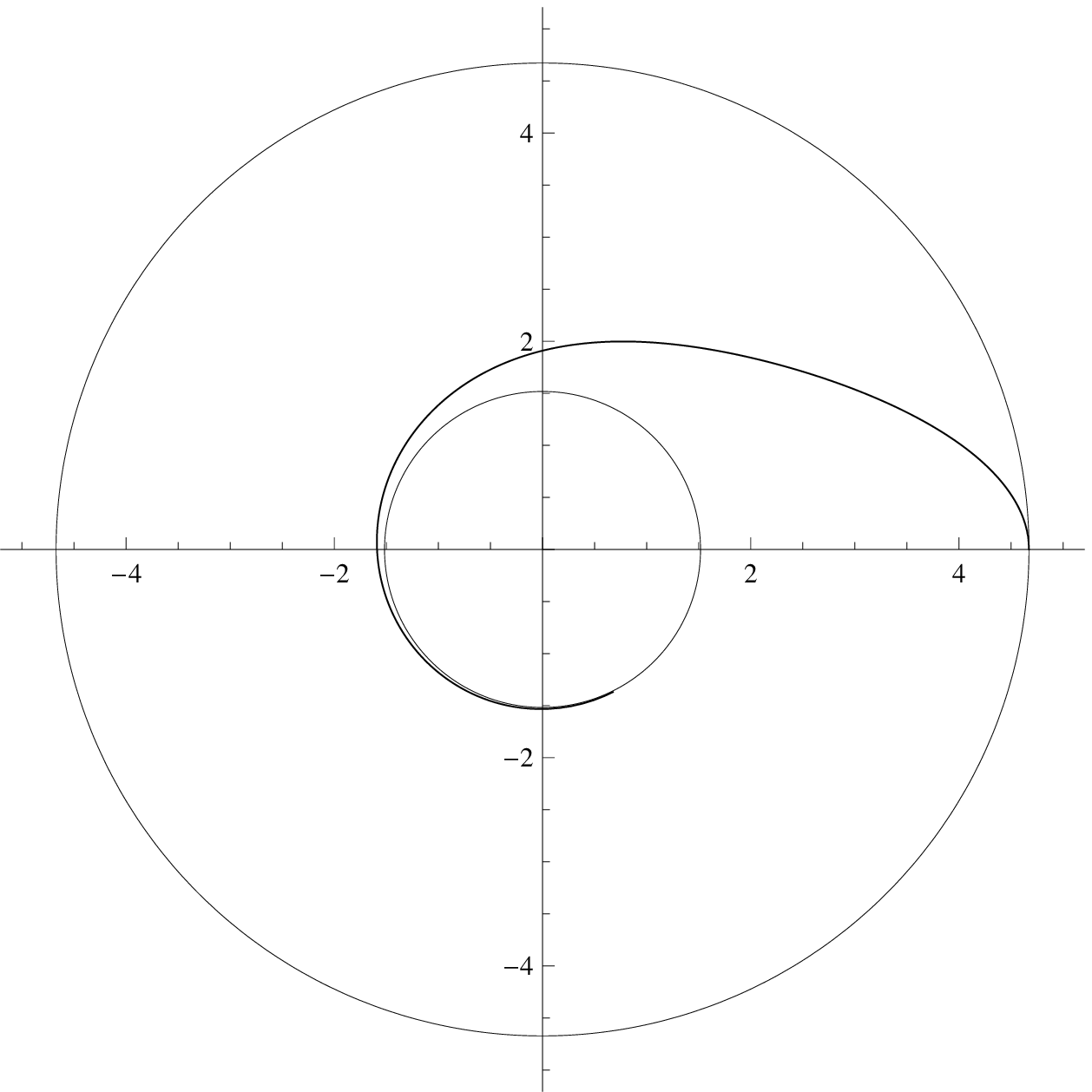}
\includegraphics[width=0.35\textwidth]{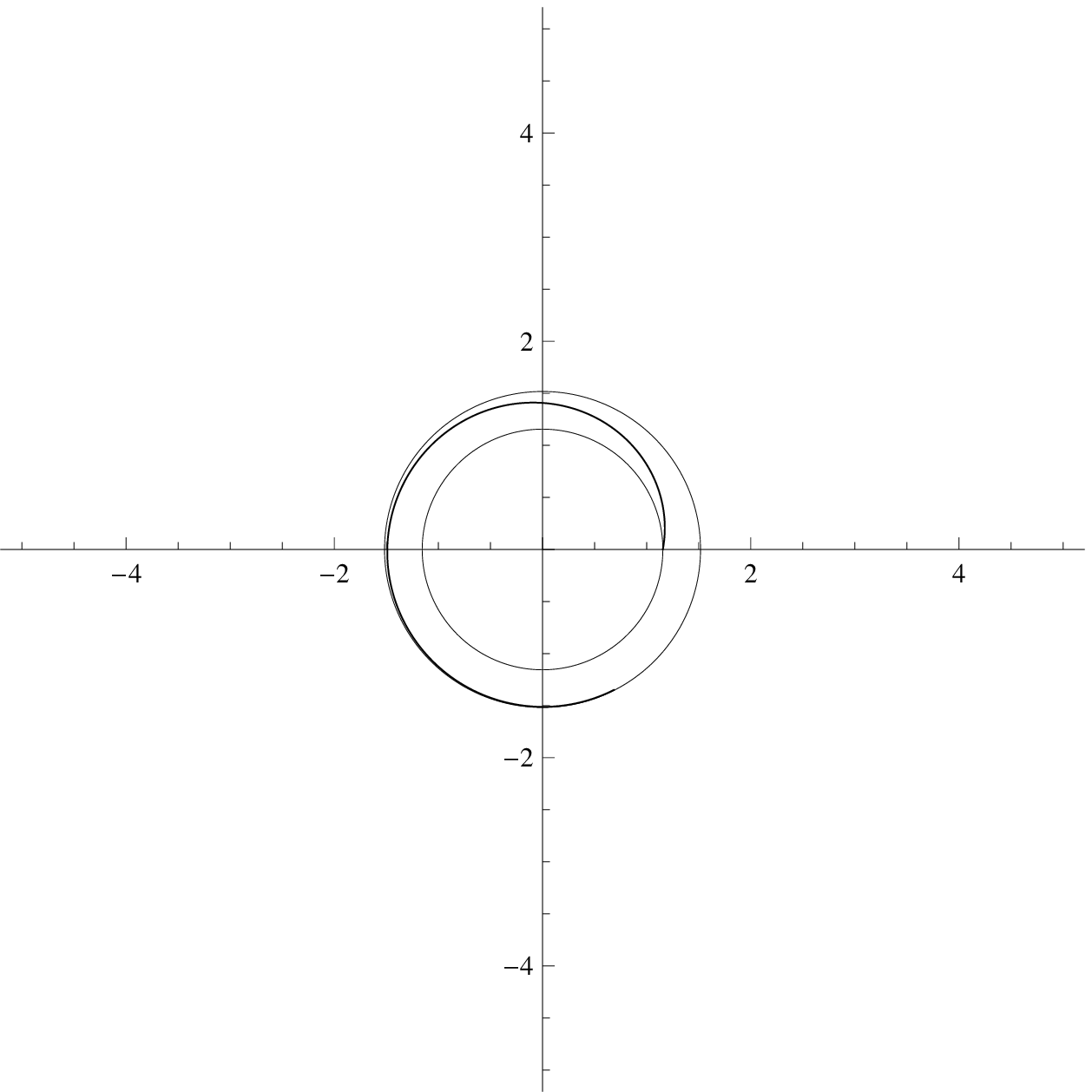}
\end{center}
\caption{The first (left figure) and second (right figure) kind of critical orbits with $L=7$, $k=1$, $\nu =-1$, $F=1$ and $G=2$.} \label{plot3}
\end{figure}
\item
The planetary orbits are constrained to oscillate between
an apoastro and a periastro. We plot in Fig. \ref{plot4}
the planetary orbit for $E=65$. We can observe that the particle completes an oscillation in  an angle less than $2\pi$ contrary to the Schwarzschild AdS black hole, where the angle is greater than $2\pi$ \cite{chandra}. 
\begin{figure}[h]
\begin{center}
\includegraphics[width=0.5\textwidth]{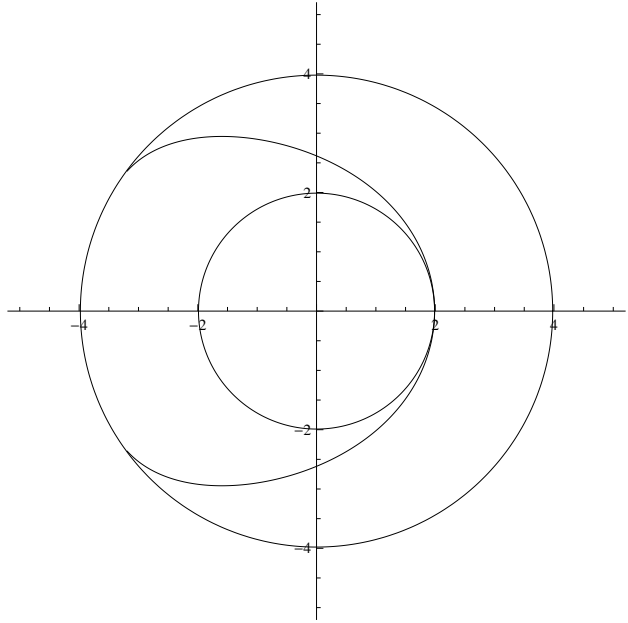}
\end{center}
\caption{The planetary orbit  with $L=7$, $k=1$, $\nu =-1$, $F=1$ and $G=2$ for $E=69.4$.} \label{plot4}
\end{figure}

\end{itemize}

It is possible to calculate the periods of the circular orbits ($r_{c.o.}$), which can be stable ($r_s$) or unstable ($r_u$) orbits using the constant of motion $\sqrt{E}$ and $L$, given by (\ref{w.11c}), which yields
\begin{equation}
T_{\tau}=\frac{2\pi r_{c.o.}(r_{c.o.}+\nu)}{L}~,
\end{equation}  
and
\begin{equation}
T_t=\frac{2\pi \sqrt{E}r_{c.o.}(r_{c.o.}+\nu)}{L f(r_{c.o.})}~,
\end{equation}
where $T_{\tau}$ is the period of the orbit with respect to the proper time and $T_t$ is the period of the orbit with respect to the coordinate time. It is worth to mention that the periods depend on the value of $\nu$ and in the limit $\nu\rightarrow 0$ these periods correspond to the periods of the circular orbits in the spacetime Schwarzschild AdS. On the other hand, for the stable circular orbits is possible to find the epicycle frequency, given by $\kappa^2=V''(r_s)/2$. 

\subsubsection{Perihelion precession}
Here, we follow the treatment performed by Cornbleet \cite{Cornbleet}, which
allows us to derive the formula for the advance of the
perihelia of planetary orbits. The starting point is to
consider the line element in unperturbed Lorentz coordinates
\begin{equation}
ds^2 = -dt^2 + dr^2 + r^2(d\theta^2 + sin^2 \theta d\phi^2)~,
\end{equation}
together with line element (\ref{metricBH}). So, considering
only the radial and time coordinates in the binomial
approximation, the transformation gives
\begin{equation}
d\tilde{t}\approx \left(1-\frac{G}{6r}+\frac{Fr^2}{2}+\frac{G\nu}{12r^2}+\frac{F\nu r}{2}\right) dt~, 
\end{equation}
\begin{equation}\label{r}
d\tilde{r}\approx \left(1+\frac{G}{6r}-\frac{Fr^2}{2}-\frac{G\nu}{12r^2}-\frac{F\nu r}{2}\right) dr~. 
\end{equation}
We will consider two elliptical orbits, one the classical
Kepler orbit in $(r, t)$ space and a hairy AdS orbit in
$(\tilde{r},\tilde{t})$ space. Then, in the Lorentz space $dA = \int_0^{\mathcal{R}}rdrd\phi=\mathcal{R}^2d\phi/2$, and hence
\begin{equation}
\frac{dA}{dt}=\frac{1}{2}\mathcal{R}^2\frac{d\phi}{dt}~,
\end{equation}
which corresponds to Kepler's second law. For the
hairy AdS case we have
\begin{equation}\label{A}
d\tilde{A}=\int_0^{\mathcal{R}}a(r)d\tilde{r}d\phi~,
\end{equation}
where $d\tilde{r}$ is given by Eq. (\ref{r}), and the binomial approximation
for the radial function $a(r)$ is
\begin{equation}
a(r)\approx r\left(1+\frac{\nu}{2r}\right)~.
\end{equation}
So, we can write (\ref{A}) as
\begin{eqnarray}\label{A2}
\nonumber d\tilde{A}&=&\int_0^{\mathcal{R}}r\left(1+\frac{\nu}{2r}\right)\left(1+\frac{G}{6r}-\frac{Fr^2}{2}-\frac{G\nu}{12r^2}-\frac{F\nu r}{2}\right) drd\phi\\
&&\approx \frac{\mathcal{R}}{2}\left(1+\frac{G}{3\mathcal{R}}-\frac{F\mathcal{R}^2}{4}+\frac{\nu}{\mathcal{R}} \right)d\phi~.
\end{eqnarray}
Therefore, applying the binomial approximation 
we obtain
\begin{eqnarray}
\nonumber \frac{ d\tilde{A}}{d\tilde{t}}&=&\frac{\mathcal{R}}{2}\left(1+\frac{G}{3\mathcal{R}}-\frac{F\mathcal{R}^2}{4}+\frac{\nu}{\mathcal{R}} \right)\frac{d\phi}{d\tilde{t}}\\
&&\approx \frac{\mathcal{R}}{2}\left(1+\frac{G}{3\mathcal{R}}-\frac{F\mathcal{R}^2}{4}+\frac{\nu}{\mathcal{R}} \right) \left(1+\frac{G}{6\mathcal{R}}-\frac{F\mathcal{R}^2}{2}-\frac{G\nu}{12\mathcal{R}^2}-\frac{F\nu \mathcal{R}}{2}\right) \frac{d\phi}{dt}~.
\end{eqnarray}
So, using this increase to improve the elemental angle
from $d\phi$ to $d\tilde{\phi}$. Then, for a single orbit
\begin{equation}\label{so}
\int_0^{\Delta\tilde{\phi}}d\tilde{\phi}=\int_0^{\Delta\phi=2\pi}\left(1+\frac{G}{2\mathcal{R}}-\frac{3F\mathcal{R}^2}{4} + \frac{\nu}{\mathcal{R}} \right)d\phi~,
\end{equation}
where we have neglected products of $G$, $F$ and $\nu$.
The polar form of an ellipse is given by
\begin{equation}\label{ell}
\mathcal{R}=\frac{l}{1+\epsilon cos\phi},
\end{equation}
where $\epsilon$ is the eccentricity and $l$ is the semi-latus rectum.
In this way, plugging Eq. (\ref{ell}) into Eq. (\ref{so}), we obtain
\begin{eqnarray}
\Delta\tilde{\phi}=2\pi+\frac{G}{2}\int_0^{2\pi} \frac{1+\epsilon cos\phi }{l} d\phi-\frac{3F}{4}\int_0^{2\pi} \left(\frac{l}{1+\epsilon cos\phi } \right)^2 d\phi+\nu\int_0^{2\pi} \frac{1+\epsilon cos\phi }{l} d\phi~,
\end{eqnarray}
which at first order yields 
\begin{eqnarray}
\Delta\tilde{\phi}\approx 2\pi+\frac{\pi G}{l}+\frac{3\pi F l^2}{2} +\frac{2\pi\nu}{l}~.
\end{eqnarray}
Therefore, the perihelion advance has the standard value
of general relativity plus the correction term coming from cosmological constant and scalar hair. It is worth to mention that there is  a (negative) discrepancy between
the observational value of the precession of perihelion for Mercury,
$\Delta\tilde{\phi}_{Obs.}=5599.74\,(arcsec/Julian-century)$ and the total
$\Delta\tilde{\phi}_{Total}= 5603.24\,(arcsec/Julian-century)$, see \cite{Olivares:2013jza}.
Which, is possible attribute to the scalar hair correction, given $\nu= -0.359 (Km)$.

\subsection{Null geodesic}
In the next analysis, we consider two kinds of motion, for $L=0$ (radial motion), and $L>0$ (angular motion) of the photons ($m=0$).
\subsubsection{Radial motion}
\begin{figure}[h]
\begin{center}
\includegraphics[width=0.5\textwidth]{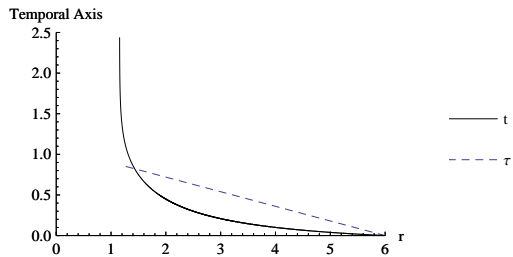}
\end{center}
\caption{The behavior of the proper ($\tau$) and coordinate ($t$) time for ingoing photons as function of $r$,  with $L=0$, $k=1$, $\nu =-1$, $F=1$ and $G=2$.} \label{plot5}
\end{figure}

In this case, the master equation (\ref{tl7}) can be written as
\begin{equation}
 \frac{dr}{d\tau}=\pm \sqrt{E}~,
\end{equation}
where ($+$) stands for outgoing photons and $(-)$ stands for ingoing photons. The solution of the above equation yields
\begin{equation}
r=\pm \sqrt{E} \tau + r_0~,
\end{equation}
where $r_0$ is an integration constant that corresponds to the initial position of the photon, as in the Schwarzschild AdS case.  The photons always fall into the horizon from an upper distance determined by the constant of motion $E=30.88$. In Fig. \ref{plot5} we plot the proper ($\tau$) and coordinate ($t$) time as function of $r$ for a photon falling from a finite distance ($r_0=6$), we can see that photons fall towards the horizon in a finite proper time. The situation is very different if we consider the trajectory in the coordinate time, where $t$ goes to infinity. 

\subsubsection{Angular motion}
\begin{figure}[h]
\begin{center}
\includegraphics[width=0.35\textwidth]{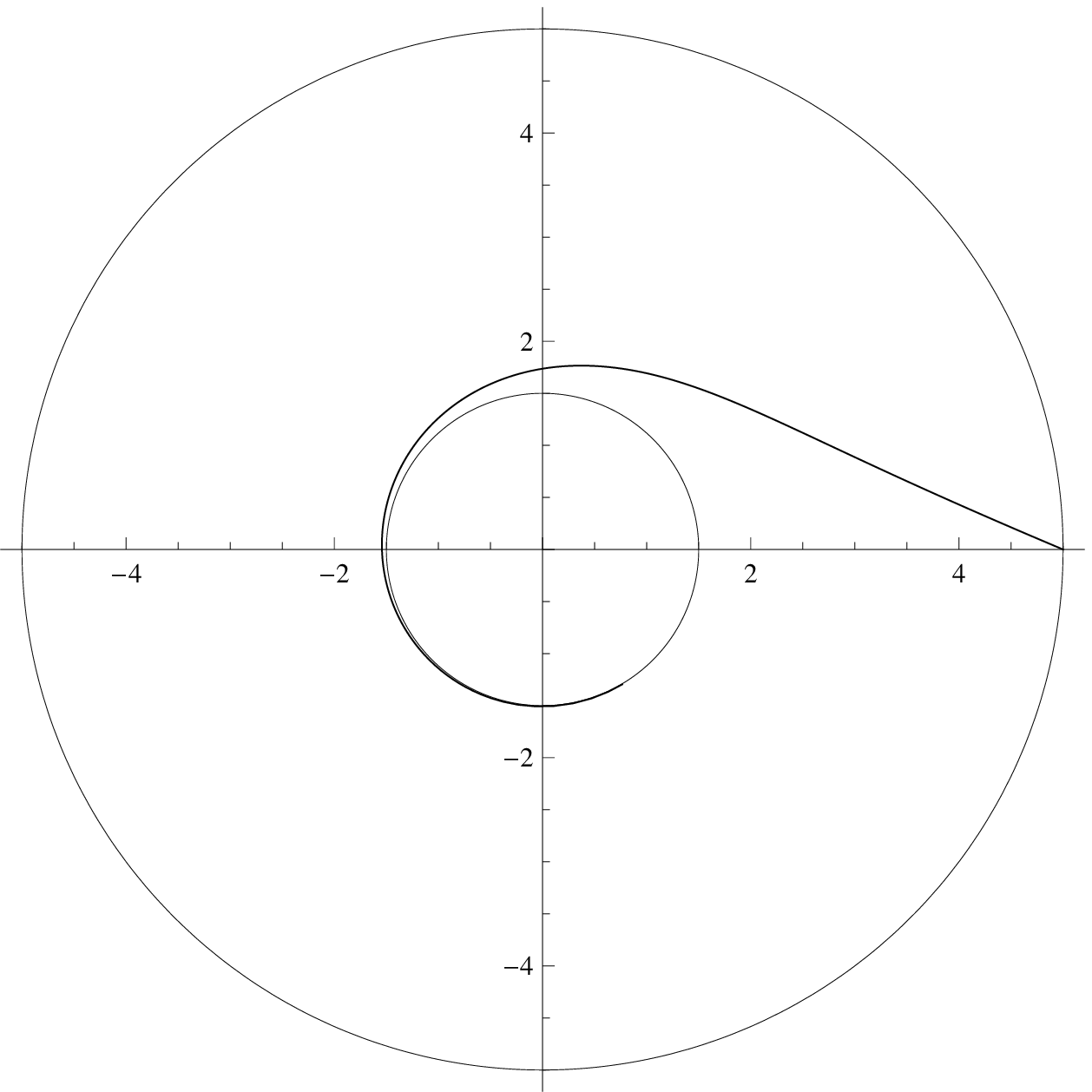}
\includegraphics[width=0.35\textwidth]{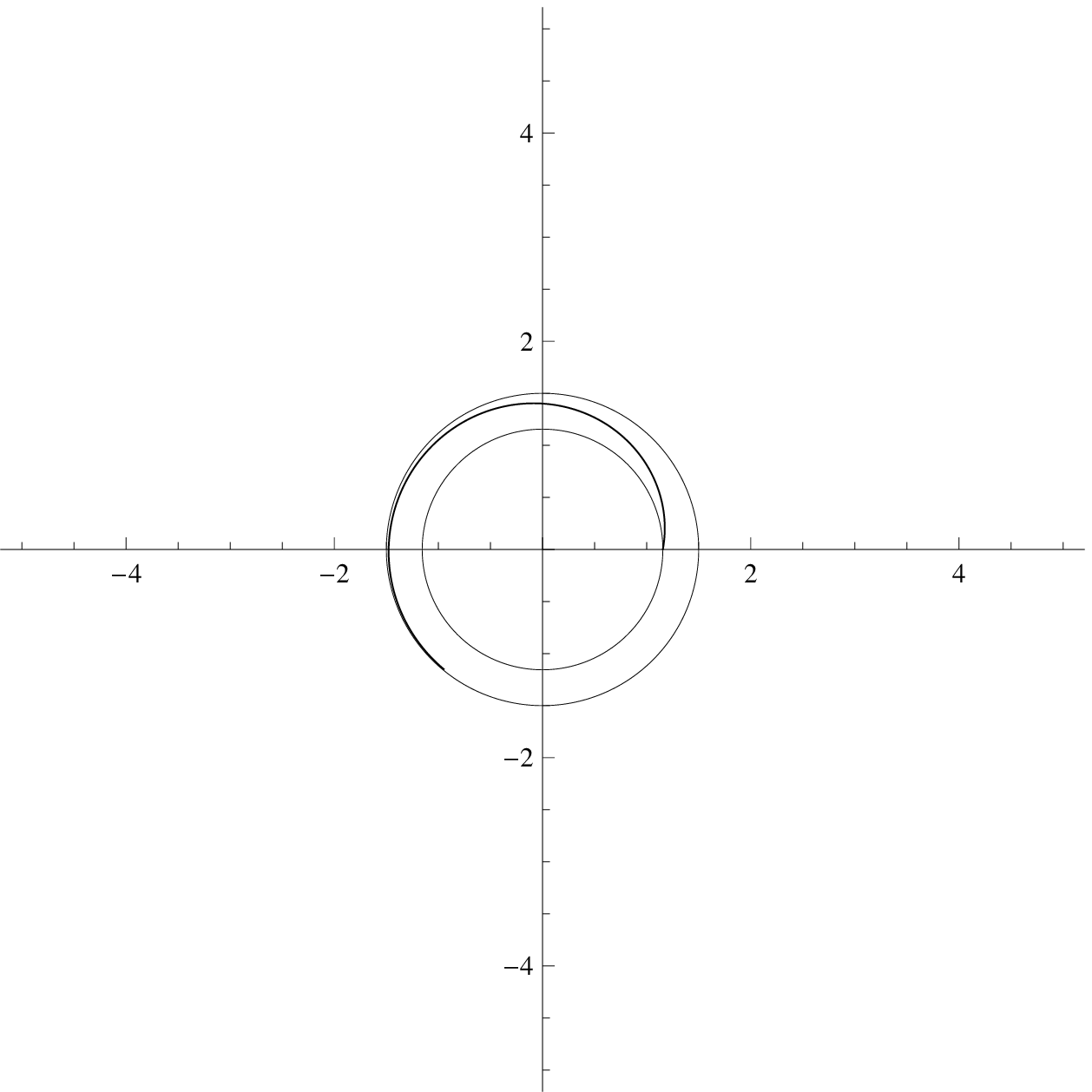}
\end{center}
\caption{The first (left figure) and second (right figure) kind of critical orbits with $L=2$ with $k=1$, $\nu =-1$, $F=1$ and $G=2$ for $E=5.58$.} \label{plot7}
\end{figure}

\begin{figure}[h]
\begin{center}
\includegraphics[width=0.5\textwidth]{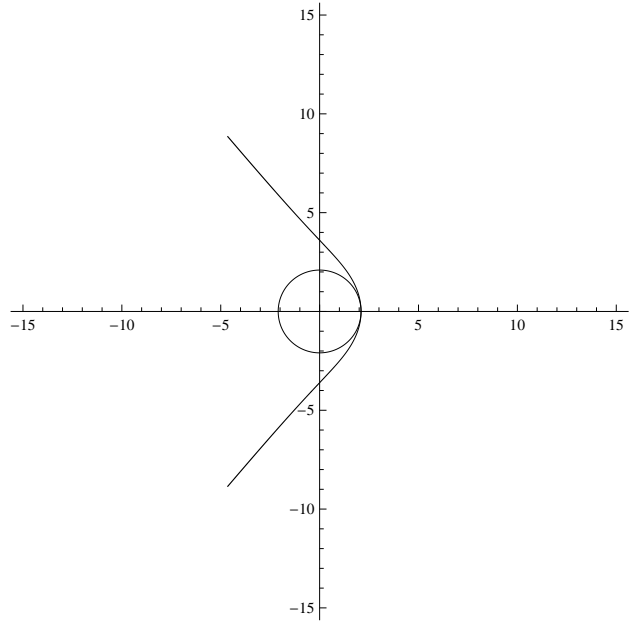}
\end{center}
\caption{The deflection of the light with $L=2$, $k=1$, $\nu =-1$, $F=1$ and $G=2$ for $E=5$.} \label{plot8}
\end{figure}

In this case, the allowed orbits for photons depend on the value of the 
impact parameter $b\equiv L/\sqrt{E}$. Next, based on the impact parameter values shown in Fig.\ref{plot6},
we give a brief qualitative description of the allowed
angular motions for photons, 
described in the following:
\begin{figure}[h]
\begin{center}
\includegraphics[width=0.5\textwidth]{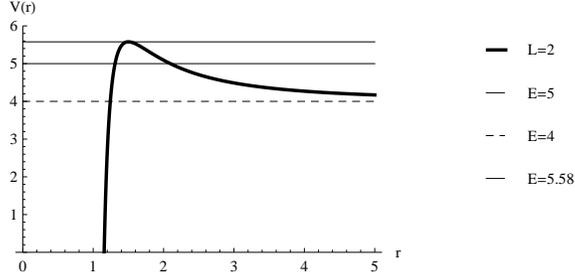}
\end{center}
\caption{The behavior of the effective potential for photons as function of $r$, with $L=2$, $k=1$, $\nu =-1$, $F=1$ and $G=2$.} \label{plot6}
\end{figure}
\begin{itemize}
  \item \emph{Capture zone}:
  If $0< b < b_{u}$, photons fall inexorably
  to the horizon,
  and its cross section,
  $\sigma$, in this geometry is \cite{wald}
  \begin{equation}\label{mr51}
    \sigma=\pi\,b_u^2
  \end{equation}
\item \emph{Critical trajectories}:
If $b=b_{u}$ ($E_u=5.58$), photons can stay in one of the unstable
  inner circular orbit of radius  $r_{u}$ ($r_u=1.5$).
  Therefore, the photons that arrive from the initial distance
  $r_i$ ($r_+ < r_i< r_u$, or $r_u< r_i<\infty$)
  can asymptotically fall into a circle of radius $r_{u}$, see Fig. \ref{plot7}.
 \item \emph{Deflection zone}. If $b_{u} <b <b_{0}=1/\sqrt{F} $, the photons can fall from the
infinity to a minimum distance $r_d=2.09$ and can come
back to the infinity. This photons are deflected, see Fig. \ref{plot8}. The
other allowed orbits correspond to photons moving into the
other side of the potential barrier, which plunges into the
singularity. In the next section, we will focus on this topic.
\end{itemize}

\subsubsection{Deflection of light}
The deflection of light is important because the deflection of light by the Sun is one of the most important test of general relativity, and the deflection of light by galaxies is the mechanism behind gravitational lenses. The distance of the closest approach $r_0$ for the metric (\ref{metricBH}) can be defined by
\begin{equation}
\left(\frac{dr}{d\phi}\right)^2=\frac{a(r)^4}{b^2}-a(r)^2f(r)~,
\end{equation}
where $b$ is the impact parameter. Now, by using the change of variables $u=1/r$, the above equation can be written as
\begin{equation}
\left(-\frac{du}{d\phi}\right)^2=\frac{1}{b^2}(1+\nu\, u)^2-(1+\nu\, u)u^2f(u)~,
\end{equation}
which at first order and applying the binomial approximation wherever necessary, we obtain
\begin{equation}\label{du}
\left(-\frac{du}{d\phi}\right)^2=\frac{1}{b^2}-F+ \frac{2\nu\, u}{b^2}-u^2+\left(\frac{G}{3}-\nu\right)u^3~.
\end{equation}
Following \cite{Shutz} we define
\begin{equation}
y\approx u- \frac{\nu}{b^2}-\frac{1}{2}\left(\frac{G}{3}-\nu\right)u^2~.
\end{equation}
So, solving for $u$ yields
\begin{equation}
u= \frac{\nu}{b^2}+y+\frac{1}{2}\left(\frac{G}{3}-\nu\right)y^2~,
\end{equation}
where we have considered first order terms. Therefore (\ref{du}) becomes 
\begin{equation}
\phi(u)= \int_{(1/b^2-F)^{-1/2}}^\infty\frac{1+(\frac{G}{3}-\nu)y}{\sqrt{\frac{1}{b^2}-F-y^2}}dy~.
\end{equation}
This can be integrated to give
\begin{equation}
\phi_\infty= {\pi\over2}+\left(\frac{G}{3}-\nu\right)\sqrt{\frac{1}{b^2}-F}~,
\end{equation}
and by considering $\Delta\phi=2\,\phi_\infty-\pi$ (see Fig. \ref{plot8}), it is possible to find
the deflection angle accurately, and it reads:
\begin{equation}
\Delta\phi= 2\left(\frac{G}{3}-\nu\right)\sqrt{\frac{1}{b^2}-F}~.
\end{equation}

Therefore, the deflection of light is given by the standard value  plus the correction 
term coming from cosmological constant ($F$) and scalar hair ($\nu$).
It is worth to mention that there is a discrepancy between the theoretical value and
the observational value of deflection light measured by Eddington and Dyson in the solar 
eclipse of March 29, 1919. For Sobral expedition this value is
$\Delta{\phi}_{Obs.}= 1.98 \pm 0.16''$ and $\Delta{\phi}_{Obs.}= 1.61 \pm 0.40''$ for 
the Principe expedition \cite{Straumann}. Currently, the mean value 
is $\Delta{\phi}_{Obs.}= 1.89$ \cite{Huang:2013jqa}. 
So, by attributing this discrepancy to the scalar hair correction and neglect the contribution 
from cosmological constant, gives $\nu= -0.386\, (Km)$, for $\Delta{\phi}_{Obs.}= 1.98$, and  
$\nu= -0.235\, (Km)$, for the mean value of $\Delta{\phi}_{Obs.}$

\subsubsection{Gravitational time delay}
An interesting relativistic effect in the propagation of light rays is the apparent delay in the time of propagation for a light signal passing near the Sun, which is a relevant correction for astronomic observations, and is called the Shapiro time delay. The time delay of Radar Echoes corresponds to the determination of the time delay of radar signals which are transmitted from the Earth through a region near the Sun to another planet or  spacecraft and then reflected back to the Earth. The time interval between emission and return of a pulse as measured by a clock on the Earth is
\begin{equation}
t_{12}=2\, t(r_1,\rho_0)+2\, t(r_2,\rho_0)~,
\end{equation}
where $\rho_0$ is the closest approach to the Sun. Now, in order to calculate the time delay we use (\ref{tl7}) and the coordinate time
\begin{equation}
\dot{r}=\dot{t}\,\frac{dr}{dt}=\frac{\sqrt{E}}{f(r)}\frac{dr}{dt}~.
\end{equation}  
So, (\ref{tl7}) can be written as
\begin{equation}\label{ct}
\frac{\sqrt{E}}{f(r)}\frac{dr}{dt}=\sqrt{E-\frac{L^2}{a(r)^2}f(r)}~.
\end{equation}
By considering $\rho_0$ the closest approach to the Sun, $dr/dt$ vanishes, so that
\begin{equation}\label{TD1}
\frac{E}{L^2}=\frac{f(\rho_0)}{a(\rho_0)^2}~.
\end{equation}
Now, by inserting (\ref{TD1}) in (\ref{ct}), the coordinate time which the light requires to go from $\rho_0$ to $r$ is
\begin{equation}
t(r,\rho_0)=\int_{\rho_0}^r \frac{dr}{f(r)\sqrt{1-\frac{a(\rho_0)^2}{f(\rho_0)}\frac{f(r)}{a(r)^2}}}~.
\end{equation}
So, at first order correction we obtain
\begin{eqnarray}
\nonumber t(r, \rho_0)&=&\sqrt{r^2-\rho_0^2}+\frac{G}{3}ln\left(\frac{r+\sqrt{r^2-\rho_0^2}}{\rho_0}\right)\\
&&+\frac{1}{2}\left(\frac{G}{3}+\nu\right)\sqrt{\frac{r-\rho_0}{r+\rho_0}}-\frac{F}{6}\sqrt{r^2-\rho_0^2}\left(2r^2+\rho_0^2\right)~.
\end{eqnarray}
Therefore, for the circuit from point 1 to point 2 and back, the delay in the coordinate time is
\begin{equation}
\Delta t := 2\left[t(r_1, \rho_0)+t(r_2,\rho_0)-\sqrt{r_1^2-\rho_0^2}-\sqrt{r_2^2-\rho_0^2}\right]=t_G+t_F+t_{\nu}~,
\end{equation}
where
\begin{eqnarray}
t_G&=&\frac{G}{3}\left[ 2\, ln\left(\frac{(r_1+\sqrt{r_1^2-\rho_0^2})(r_2+\sqrt{r_2^2-\rho_0^2})}{\rho_0^2}\right)+\sqrt{\frac{r_1-\rho_0}{r_1+\rho_0}}+\sqrt{\frac{r_2-\rho_0}{r_2+\rho_0}}\right]~,\\
t_F&=&-\frac{F}{6}\left[\sqrt{r_1^2-\rho_0^2}\left(2r_1^2+\rho_0^2\right)+\sqrt{r_2^2-\rho_0^2}\left(2r_2^2+\rho_0^2\right)\right]~,\\
t_\nu&=&\nu\left(\sqrt{\frac{r_1-\rho_0}{r_1+\rho_0}}+\sqrt{\frac{r_2-\rho_0}{r_2+\rho_0}}\right)~.
\end{eqnarray}
For a round trip in the solar system, we have ($\rho_0<<r_1,r_2$)
\begin{equation}
\Delta t \approx \frac{2G}{3}\, ln \left(\frac{4r_1r_2}{\rho_0}\right)+2\left(\frac{G}{3}+\nu\right)-\frac{F}{3}\left(r_1^3+r_2^3\right)~.
\end{equation}
Therefore, as in the previous cases the time delay has the standard value
of general relativity plus the correction term coming from 
cosmological constant and scalar hair.

\section{Concluding comments}
\label{sec:conclution}

We have considered a four-dimensional asymptotically AdS black hole with scalar 
hair \cite{Gonzalez:2013aca}. These solutions asymptotically give the Schwarzschild anti-de
Sitter solution, and they are characterized by a scalar field with a
logarithmic behavior, being regular everywhere outside the event
horizon and null at spatial infinity,  and by a self-interacting
potential, which tends to the cosmological constant at spatial
infinity. The equations for the geodesics were solved numerically in order to study their behavior.
We note that radial motion results to be equivalent to the  Schwarzschild AdS space-time \cite{chandra}. 
 Mainly, we have found that it is possible to find bounded 
orbits like planetary orbits in the background of a four-dimensional asymptotically AdS black holes with scalar 
hair. However, the periods associated to circular orbits 
are modified by the presence of the scalar hair. Besides, we have found that some classical 
tests such as perihelion precession, deflection of light and gravitational time 
delay have the standard value of general relativity plus a correction term coming 
from the cosmological constant and scalar hair. Finally, we found a specific value of the parameter associated to the scalar hair, in order to explain the discrepancy between the theory and the observations, for the perihelion precession of Mercury ($\nu= -0.359 (Km)$) and light deflection ($\nu= -0.386\, (Km)$ for $\Delta{\phi}_{Obs.}= 1.98$). Interestingly, these values are of the same order and sign. In furthering our understanding, it would be interesting to study the motion 
of massless and massive particles in a charged hairy black hole. Work in this direction is in progress.

\acknowledgments 

This work was funded by Comisi\'{o}n
Nacional de Ciencias y Tecnolog\'{i}a through FONDECYT Grants 11140674 (PAG) and 11121148 (YV). M. O. thanks to PUCV.

\end{document}